\PassOptionsToPackage{hyphens}{url}
\documentclass[epj,twocolumn]{webofc}
\usepackage[varg]{txfonts}   % Web of Conferences font
\usepackage{epstopdf}
\usepackage{physmacros}

\woctitle{LHCP 2013}

\begin{document}

\title{The ATLAS Tile Calorimeter Calibration and Performance}

\author{Christopher Meyer\inst{1}\fnsep\thanks{\email{chris.meyer@cern.ch}} on behalf of the ATLAS Collaboration}

\institute{The Enrico Fermi Institute, The University of Chicago}

\abstract{A brief summary of the hadronic calorimeter calibration systems and performance results, in the ATLAS detector at the LHC is given.}

\maketitle

%\linenumbers

\section{Introduction}
\label{intro}
\begin{figure}
\centering
\includegraphics[width=4.6cm,clip]{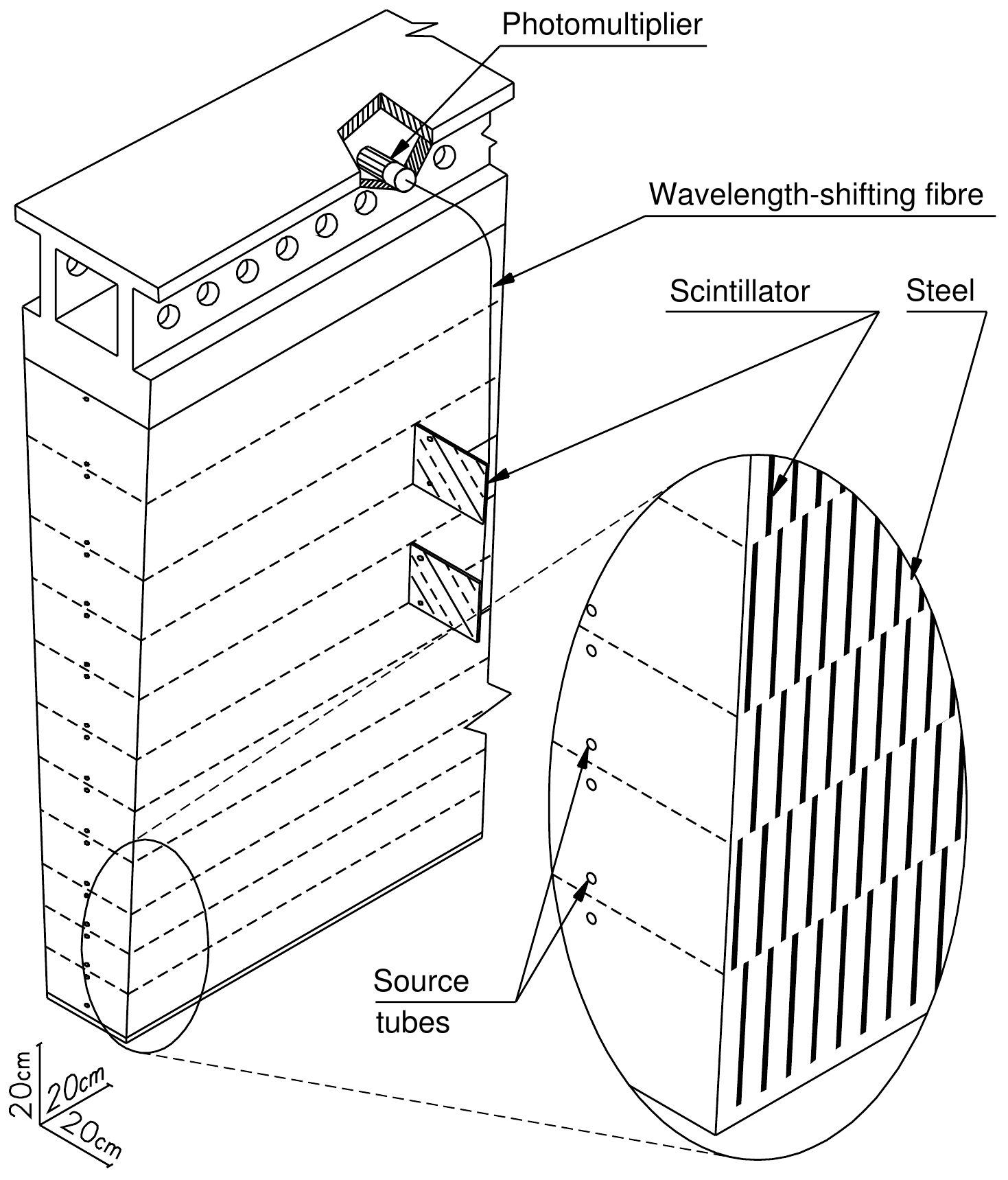}
\caption{One module of the tile calorimeter, showing alternating steel and scintillating material. \cite{Aad:atlaslist}}
\label{fig:tilecutout}
\end{figure}

\begin{figure}
\centering
\includegraphics[width=6cm,clip]{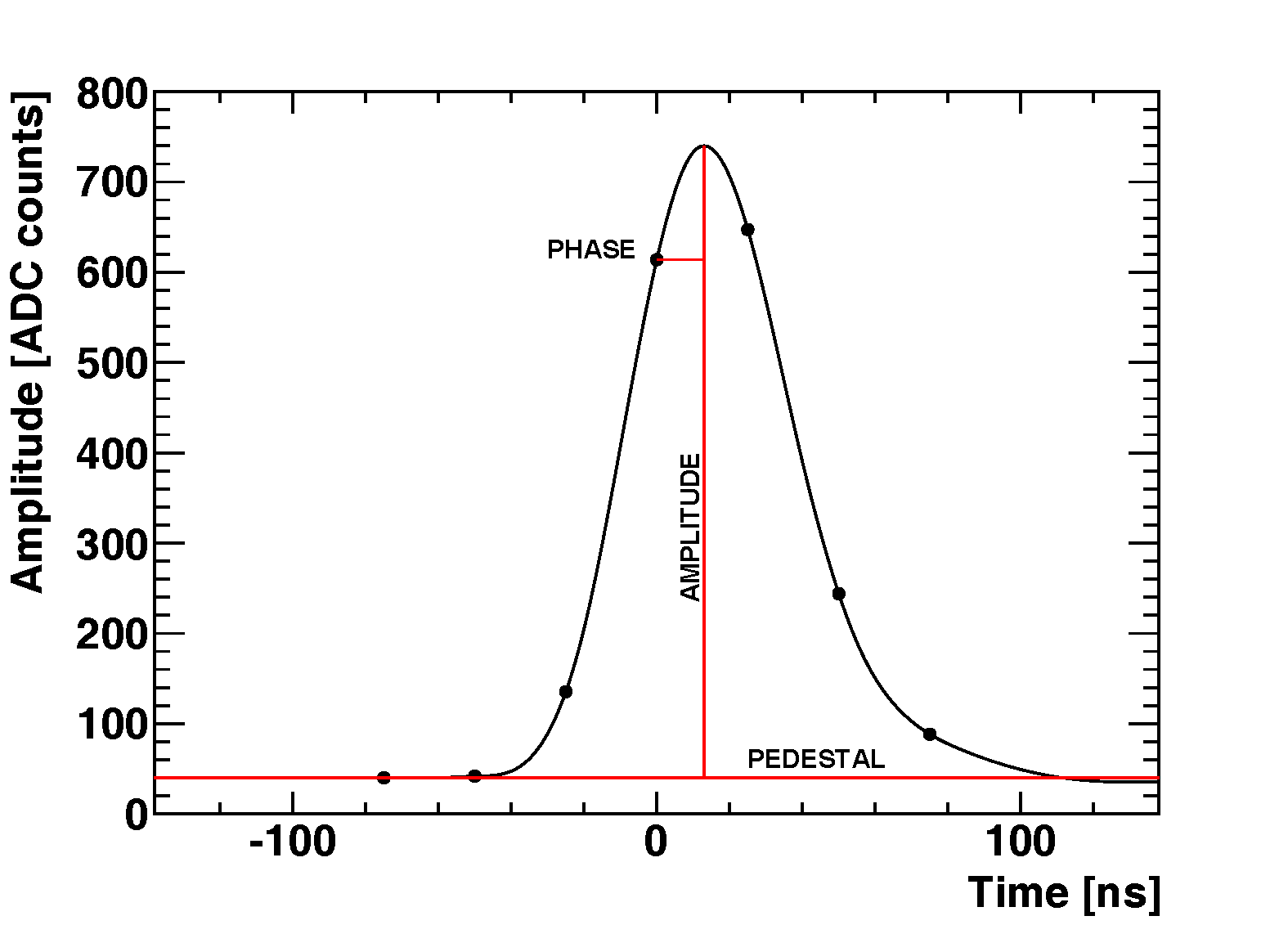}
\caption{Example pulse shape showing the 7 sampled points, the fitted pulse shape, and the amplitude, timing, and pedestal components.}
\label{fig:pulse}
\end{figure}

The ATLAS \cite{Aad:2008zzm} tile calorimeter (TileCal) \cite{Aad:2010af} is a sampling, hadron calorimeter, located at the LHC.
The central barrel portion covers $|\eta| < 0.8$, while the extended partitions on either side cover out to $|\eta| < 1.7$.
It is composed of alternating layers of plastic scintillating material and steel (see figure \ref{fig:tilecutout}), grouped to create cells.
In total, 9852 photmultiplier tubes (PMTs) read out energy deposited in the detector.
The calorimeter was designed to have a resolution of $\sigma/E = 50\%/\sqrt{E} \oplus 3\%$, and enable a jet energy calibration uncertainty of $<1\%$.
The PMT signal is first shaped into a pulse with FWHM$\sim 50$ ns, then sampled by a 40 MHz analog to digital converter (ADC).
To cover the full dynamic range two gain channels are used, so that smaller signals are amplified before being sampled by the ADC.
This results in precise signal reconstruction over the full dynamic range.
The resulting seven samples of the pulse are fit using the optimal filter method \cite{Usai:2011zz} to determine the amplitude in ADC counts and timing in ns (see figure \ref{fig:pulse}).
Although the digital signal processor which performs the fit has limited resolution due to the use of fast lookup tables, the majority of the signals produced by physics show negligible difference when compared to the full offline reconstruction.

To convert from ADC counts to energy in MeV a series of calibrations is applied:
\begin{itemize}
\item \textbf{Charge injection system (CIS)}:
provides a calibration $C_\mathrm{CIS}$ from ADC counts to pC.
\item \textbf{Test beam}:
the initial calibration $C_\mathrm{test beam}$ converting pC to \MeV, derived using test beam results.
\item \textbf{Cesium (\Cs)}:
provides a relative calibration $C_\mathrm{Cs}$ to account for changes in the scintillating material, optical fibers, and PMTs since deriving the test beam calibration \cite{Adragna:2009zz}.
\item \textbf{Laser}:
provides a relative calibration $C_\mathrm{laser}$ to account for the drift of the PMTs and optical fibers between \Cs runs.
\end{itemize}
The total calibration applied to the fitted pulse amplitude $A$ in ADC counts to derive the measured electromagnetic scale energy $E$ is:
%\begin{linenomath}
\begin{equation*}
E[\MeV] = C_\mathrm{test beam} \times C_\mathrm{Cs} \times C_\mathrm{laser} \times C_\mathrm{CIS} \times A[\mathrm{ADC}]
\end{equation*}
%\end{linenomath}
Below, the performance of the calibration systems and TileCal as a whole are discussed in more detail.

\section{Calibration Systems}
\label{calsys}
\begin{figure}
\centering
\includegraphics[width=6.0cm,clip]{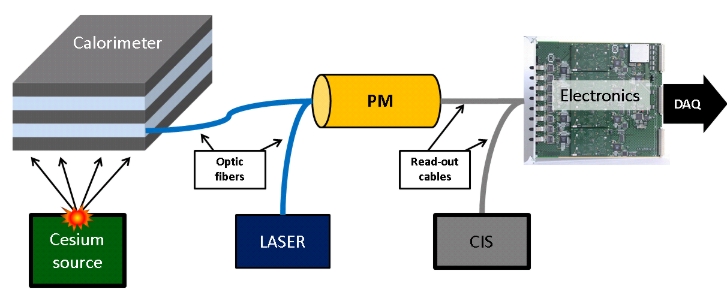}
\caption{Overview of the calibration systems in TileCal, and which portions of the readout electronics they are sensitive to.}
\label{fig:calib}
\end{figure}

In addition to providing up to date calibration constants for physics signals, the calibration systems are also staggered such that issues throughout the physics readout path can be diagnosed.
For example, if an issue is found in the calibration data from the \Cs and Laser systems, but not in CIS, the problem likely lies in the PMT (see figure \ref{fig:calib}).
This setup is particularly useful for trouble-shooting unstable high-voltage power supplies and  pathological PMTs.
For this reason, as well as providing high quality calibrations, it is important to keep track of performance for the various calibration systems, as outlined below.

\subsection{Cesium}
\label{cs}
The cesium system \cite{Aad:2010af,Starchenko:2002ju,Shalanda:2003rq} circulates a radioactive \Cs source through 10 km of tubes in TileCal.
Photons with energy of $0.662\MeV$ are emitted at a known rate, and the integrated current is read out from each cell as the source passes by.
This provides a relative calibration $C_\mathrm{Cs}$, which can be combined with the preliminary test beam calibration factor $C_\mathrm{test beam}$ to convert pC to MeV.
Three \Cs sources are circulated through TileCal approximately once per month to measure the changing conditions of the scintillating material (resulting from irradiation by physics runs) as well as the PMTs (changes of the PMT vacuum due to signals flushing out impurities).
The results have been cross checked in situ using muons from cosmic rays as well as physics runs.
By using tracking information to determine the muon momentum, the expected energy deposited in the calorimeter can be compared with the actual energy measured.

\subsection{Laser}
\label{laser}
\begin{figure}
\centering
\includegraphics[width=4.8cm,clip]{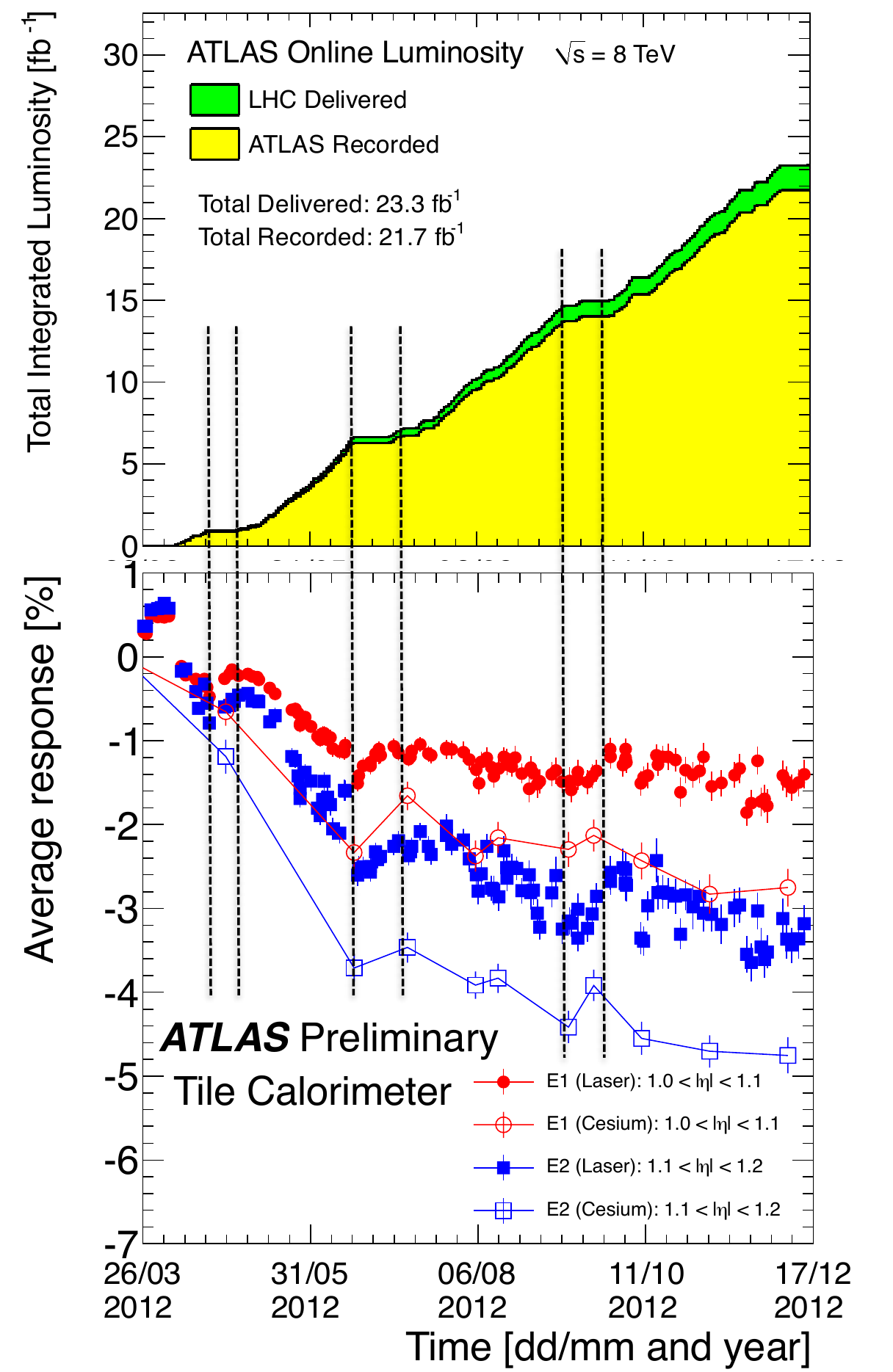}
\caption{Evolution of the \Cs and laser calibration constants during 2012 data taking. Shown for the E1 and E2 cells, located in the crack between the long and extended barrel of TileCal \cite{Aad:tilepub1}.}
\label{fig:laserces}
\end{figure}

\begin{figure}
\centering
\includegraphics[width=4.8cm,clip]{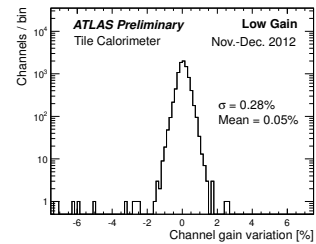}
\caption{Average variation of the laser calibration for low gain channels in TileCal, shown for the last two months of 2012 data taking \cite{Aad:tilepub1}.}
\label{fig:laser}
\end{figure}

The laser system \cite{Aad:2010af,Viret:2010zz} sends a pulse of light with a known amplitude through fiber-optic cables to each PMT.
This provides a relative calibration $C_\mathrm{laser}$ with respect to the cesium, tracking the PMT drift over the period of a month (the time interval between \Cs runs).
By definition $C_\mathrm{laser}=1$ immediately following a \Cs run.
The evolution of the laser and \Cs calibrations are compared in figure \ref{fig:laserces} for E1 and E2 cells, located in the crack between the long and extended barrels.
Because these cells have the most direct exposure to the interaction point, they are expected to show the largest change.
Differences are visible because the laser system is only sensitive to the drift of the PMTs, while the \Cs calibration is also affected by changes in the scintillating material.
For the majority of the cells the PMT and optical fiber response is much more stable, as shown in figure \ref{fig:laser}.
During the last two months of 2012 running, the laser calibration changed only $0.5\%$ on average.

\subsection{Charge Injection System}
\label{cis}
\begin{figure}
\centering
\includegraphics[width=4.8cm,clip]{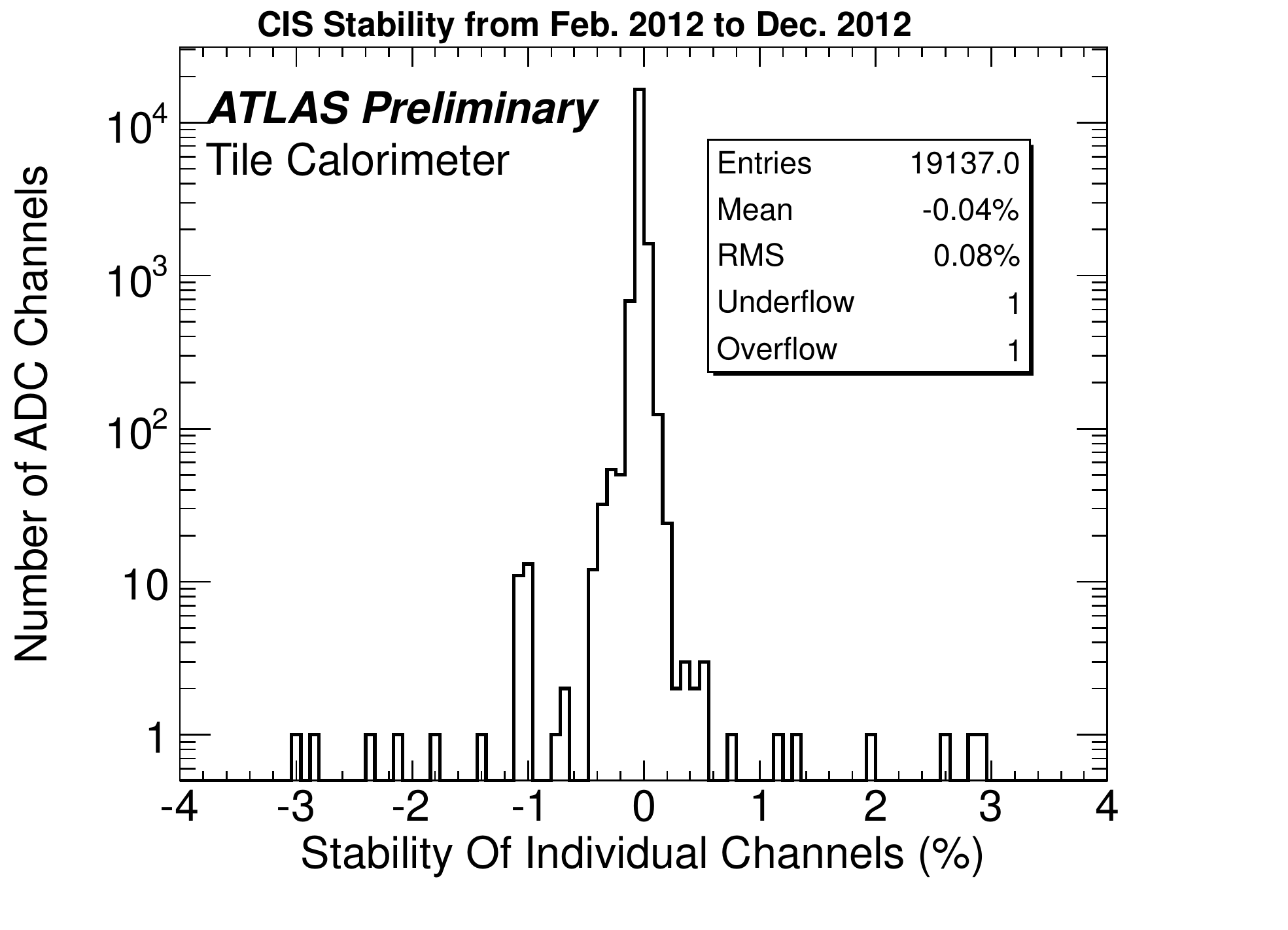}
\caption{Distribution of fractional change of CIS constants over the 2012 run period \cite{Aad:tilepub1}.}
\label{fig:cis}
\end{figure}

The charge injection system \cite{Aad:2010af,Anderson:2005ym} injects a pulse of known charge and records the output in ADC counts.
% This result is averaged over 60 injected pulses, then repeated for a range of input charges.
After scanning a range of input charge, the results are used to derive the $C_\mathrm{CIS}$ conversion factor, converting ADC counts to pC.
CIS is generally very stable, with an average drift of $0.4\%$ during 2012 running (see figure \ref{fig:cis}).
While instabilities in the electronics can cause shifts of the individual channel calibrations of up to 1\%, these jumps are rare and are generally corrected within 28 days.

\section{Performance}
\label{perf}

\begin{figure}
\centering
\includegraphics[width=4.8cm,clip]{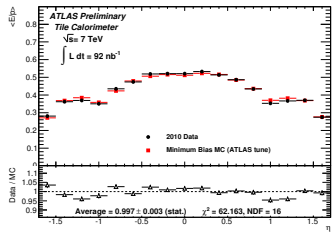}
\caption{Average energy (E) over momentum (p) of single particles measured in data (black) and MC simulation (red) \cite{Aad:tilepub2}.}
\label{fig:eoverp}
\end{figure}

\begin{figure}
\centering
\includegraphics[width=4.8cm,clip]{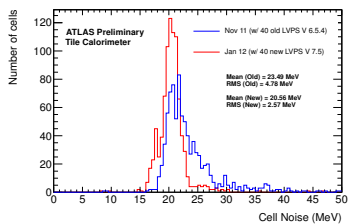}
\caption{Cell noise in TileCal readout electronics, comparing the new (red) low voltage power supplies with the old (blue) \cite{Aad:tilepub2}.}
\label{fig:noise}
\end{figure}

In general TileCal provided stable, good quality data throughout Run I at the LHC.
The readout electronics continue to perform well, as seen by the comparing of a pulse shape from the readout electronics with the reference pulse shape.% in figure \ref{fig:pulseshape}.
The calibration systems obtain stable results, and correct for any changes due to irradiation before it can affect physics.
Good agreement is observed between data and MC simulation when plotting the mean $E/p$ for single particle response, shown in figure \ref{fig:eoverp}.
The high-voltage power supplies also performed very well, with minimal deviations from the set value.
During long shutdown 1 new radiation-hard low-voltage power supplies are being installed, which also provide more Gaussian shaped noise with a smaller RMS (see figure \ref{fig:noise}).
This will reduce the number of tripped modules during physics runs, as well as further improve the resolution for physics analyses which rely on TileCal.

% \begin{figure}
% \centering
% \includegraphics[width=4.8cm,clip]{figures/PS_stdDev_hi}
% \caption{Pulse shape measured in situ compared with the reference pulse shape, for TileCal readout electronics \cite{Aad:tilepub2}.}
% \label{fig:pulseshape}
% \end{figure}

\section{Summary}
The calibration systems (\Cs, Laser, and CIS) provide an up-to-date status of TileCal performance.
In general, the calibrations have shown to be stable over time. 
When drifts arise due to changes in the scintillating material, optical fibers, PMTs, or readout electronics they are quickly caught and corrected for.
TileCal has consistently provided good quality, well calibrated data for the duration of Run I at the LHC.

%\end{linenumbers}

\bibliography{LHCP2013-Meyer-Tile}

\end{document}